\documentstyle[aps,fleqn,epsfig]{revtex}
\title{Noise in refrigerating tunnel junctions and in microbolometers}
\author{D. V. Anghel\thanks{Permanent address: NIPNE  -- 
``Horia Hulubei'', P.O.BOX MG-6, R.O.-76900 
Bucure\c sti - M\u agurele, Romania.} and J. P. Pekola\\
Department of Physics, University of Jyv\"askyl\"a,\\
P.O. Box 35 (Y5), 40351 Jyv\"askyl\"a, Finland}

\begin{document}
\maketitle

\begin{abstract}
Microrefrigerators based on normal metal--insulator--superconductor 
(NIS) junctions represent a very attractive 
alternative to cool the microbolometers and calorimeters 
for astrophysical observations in space-borne experiments. The 
performance in such measurements requires a good knowledge of 
the noise sources in the detectors. In this paper we present 
detailed calculations of the thermal fluctuations and of the 
noise equivalent power due to the heat transfer through the 
NIS junctions or due to the thermal contact between different 
subsystems of the detector. The influence of the background 
radiation will also be evaluated. Analytical approximations, 
valid at low temperatures, are given.
\end{abstract}

\section{Introduction} \label{intro}

The most sensitive detectors, suitable for observation of cosmic 
sources in far-infrared and X-ray bands, are the cryogenic 
bolometers 
and microcalorimeters. These detectors operate typically at a 
temperature of 
about 100 mK. An attractive method for the last stage of detector 
cooling 
(from 300 mK to 100 mK) in space-borne experiments is based on 
normal metal--insulator--superconductor (NIS) refrigerating 
junctions 
\cite{jukka,trap}. The sensitivity required by such experiments 
motivates the study of the noise sources in the detectors. 

A thermal detector system, such as a bolometer or a calorimeter, 
consists of a thermal sensing element (TSE) which is connected to a 
heat 
sink. The TSE typically consists of an absorber 
and a thermometer. The thermometer can be a transition-edge sensor 
\cite{irwin1,lee,luukanen2}, or a (SI)NIS tunnel junction 
thermometer \cite{nahum1,nahum2}. In the rest of this paper we 
shall study the temperature fluctuations and the noise equivalent 
power ($NEP$) in the electron system (ES) -- which is the actual 
sensing device -- of the TSE. In the presence of the optical load 
due to the background radiation, in the concrete calculations, the 
aim will be to keep the temperature of the ES, $T_{\rm e}$, around 
0.1 K. The temperature of the heat bath, $T_2$, could be 
significantly higher then $T_{\rm e}$, say $T_2\approx 0.3$ 
K, in which case the refrigerating junctions are connected directly 
to the TSE (the {\em direct cooling method}), or $T_2$ could be 
about the same as $T_{\em e}$, say $T_2 \approx T_{\rm e} \approx 
0.1$ K, when a refrigerating system is used to cool the heat bath 
(the {\em indirect cooling method}) \cite{leivo_t,leivo} and, 
eventually, another set of junctions connected  
to the TSE is used to compensate for the optical load (this 
will be called the {\em combined cooling method}). 

Let us start by supposing that a set of NIS cooling junctions is 
connected to the TSE and calculate the tunneling currents, the 
power extracted from the ES of the TSE, and the shot noise in 
the tunneling current. Following Ref. \cite{jochum} we define four 
tunneling currents between the ES of the normal TSE and the 
superconductor (we assume that the tunnel barrier is sufficiently 
thick, so we neglect the Andreev reflection of electrons between the 
normal metal and the superconductor \cite{averin}):
\begin{eqnarray}\begin{array}{lcl}
j_1(\epsilon)$=$g(\epsilon)f_{\rm e}(\epsilon-eV,T_{\rm e})[1-
f_{\rm s}(\epsilon,T_{\rm s})]/e^2R_{\rm T}, \\
j_2(\epsilon)$=$g(\epsilon)f_{\rm e}(\epsilon+eV,T_{\rm e})[1-
f_{\rm s}(\epsilon,T_{\rm s})]/e^2R_{\rm T}, \\
j_3(\epsilon)$=$g(\epsilon)[1-f_{\rm e}(\epsilon-eV,T_{\rm e})]
f_{\rm s}(\epsilon,T_{\rm s})/e^2R_{\rm T}, \\
j_4(\epsilon)$=$g(\epsilon)[1-f_{\rm e}(\epsilon+eV,T_{\rm e})]
f_{\rm s}(\epsilon,T_{\rm s})/e^2R_{\rm T},
\end{array} \label{curenti}
\end{eqnarray}
where $f_{\rm e,s}(\epsilon,T_{\rm e,s})$ represent the populations 
of the 
electron (in the normal metal) and quasiparticle (in the 
superconductor) 
energy levels ($T_{\rm s}$ is an effective temperature in the 
superconductor, used to describe the population of the quasiparticle 
energy levels). $V$ is the voltage 
across the junction, $e$ is the elementary charge, and 
$g(\epsilon)=\theta(|\epsilon|-\Delta) |\epsilon| / 
\sqrt{\epsilon^2-\Delta^2}$ is 
the normalized density of states in the superconductor, where 
$\theta$ is the Heaviside step function. 
The energy $\epsilon$ 
is measured from the Fermi energy in the superconductor 
and in Eqs. (\ref{curenti}) it is always taken in absolute value. 
Although $f_{\rm e,s}$ may not be Fermi distributions in our case
of nonequilibrium \cite{gueron} we make the assumption that 
$1-f_{\rm e}(-\epsilon,T_{\rm e})=f_{\rm e}(\epsilon,T_{\rm e})$ 
(which is an identity for a Fermi distribution) to transform the 
expressions that involved negative $\epsilon$. In what follows we 
shall concentrate on the 
junction where the flux of electrons is oriented from the normal 
metal into the superconductor, where $eV$ is 
positive. Using the definitions given in Eqs. (\ref{curenti}), 
the particle and excitation fluxes, 
$\dot{N}_{\rm J}$ and $\dot{N'}_{\rm J}$, respectively, can be written as 
\cite{jochum}
\begin{eqnarray}
\dot{N}_{\rm J} &=& \frac{1}{e} I_{\rm e} = 
\int_\Delta^\infty (j_1-j_2-j_3+j_4)\, d\epsilon 
\label{charge_c} \\
&=& \frac{1}{e^2R_{\rm T}}\int_\Delta^\infty g(\epsilon)
[f_{\rm e}(\epsilon-eV,T_{\rm e})-f_{\rm e}(\epsilon+eV, 
T_{\rm e})]\, d\epsilon , \nonumber \\
\dot{N'}_{\rm J} &=& \int_\Delta^\infty (j_1+j_2-j_3-j_4)\, d\epsilon 
\label{excitation_c} \\
&=& \frac{1}{e^2R_{\rm T}}\int_\Delta^\infty g(\epsilon)
[f_{\rm e}(\epsilon-eV,T_{\rm e})+f_{\rm e}(\epsilon+eV,T_{\rm e})-
2f_{\rm s}(\epsilon,T_{\rm s})]\, d\epsilon .
\nonumber 
\end{eqnarray}
If we denote by $\epsilon_{\rm F}$ the Fermi energy of the ES in the 
TSE, then the total power extracted through a NIS junction 
can be written as 
\begin{eqnarray}
\dot{Q}'_{\rm J}&=&\int_\Delta^\infty 
[(\epsilon-eV+\epsilon_{\rm F})(j_1-j_3)+
(\epsilon_{\rm F}-\epsilon-eV)(j_4-j_2)]\, d\epsilon 
\label{power_cp} \\
&\equiv& \dot{Q}_{\rm J} + \epsilon_{\rm F} \dot{N}_{\rm J} , 
\nonumber 
\end{eqnarray}
where 
\begin{eqnarray}
\dot{Q}_{\rm J}&=&\int_\Delta^\infty [(\epsilon-eV)(j_1-j_3)-
(\epsilon+eV)(j_4-j_2)]\, d\epsilon 
\label{power_c} \\
&=& \int_\Delta^\infty [\epsilon(j_1 + j_2 - j_3 - j_4) 
- eV (j_1 - j_2 - j_3 + j_4)] \, d\epsilon . \nonumber 
\end{eqnarray}
Since the coolers consist usually of pairs of NIS junctions 
biased in opposite directions \cite{jukka}, the total power should 
be calculated as a sum of terms like the ones given in Eq. 
(\ref{power_cp}). In a stationary case the total number of 
electrons in the TSE should be constant and the terms of the type 
$\epsilon_{\rm F} \dot{N}_{\rm J}$ cancel each other and 
we are left with the sum over $\dot{Q}_{\rm J}$'s. 

The spectral density of the shot noise in the power 
extracted through the NIS junction [given by Eq. 
(\ref{power_cp})] can be written as:
\begin{eqnarray}
\langle\delta^2 \dot{Q}{'_{\rm J}}\rangle_\omega &=& 2
\int_\Delta^\infty [(\epsilon-eV+\epsilon_{\rm F})^2(j_1+j_3)+
(\epsilon_{\rm F}-\epsilon-eV)^2(j_4+j_2)]\, d\epsilon \nonumber\\
&=& \frac{2}{e^2R_{\rm T}}\int_\Delta^\infty d\epsilon
g(\epsilon)\bigg\{(\epsilon-eV)^2 \nonumber \\
& &\times [f(\epsilon-eV, T_{\rm e})(1-2f(\epsilon, T_{\rm s}))+
f(\epsilon, T_{\rm s})] \nonumber \\
& & + (\epsilon+eV)^2[f(\epsilon+eV, T_{\rm e})(1-2f(\epsilon, T_{\rm s}))+
f(\epsilon, T_{\rm s})]\bigg\} \label{pshot_1} \\
& & + \frac{2}{e^2R_{\rm T}}\epsilon^2_{\rm F}\int_\Delta^\infty 
d\epsilon
g(\epsilon)\bigg[f(\epsilon-eV, T_{\rm e})(1-2f(\epsilon, T_{\rm s})) 
\nonumber \\
& &+f(\epsilon+eV, T_{\rm e})(1-2f(\epsilon, T_{\rm s}))+
2f(\epsilon, T_{\rm s})\bigg] \label{pshot_2} \\
& & + \frac{4}{e^2R_{\rm T}}\epsilon_{\rm F}\int_\Delta^\infty 
d\epsilon
\bigg\{g(\epsilon)(\epsilon-eV) \nonumber \\
& &\times [f(\epsilon-eV, T_{\rm e})(1-2f(\epsilon, T_{\rm s}))+
f(\epsilon, T_{\rm s})] \nonumber \\
& & - (\epsilon+eV)[f(\epsilon+eV, T_{\rm e})(1-2f(\epsilon, T_{\rm s}))+
f(\epsilon, T_{\rm s})]\bigg\}  \label{pshot_3}
\end{eqnarray}
(see  \cite{kogan}, Chap. 1, for a general introduction).
In the construction of microbolometers the TSE is sufficiently 
large and the Coulomb blockade phenomenon cannot influence the 
tunneling of individual electrons. In such a case, in the calculation 
of the spectral density of the shot noise in the total power flux, the 
contributions of the terms (\ref{pshot_1}), (\ref{pshot_2}), and 
(\ref{pshot_3}), corresponding to different junctions are simply added 
to each other. As a result, one would obtain a fluctuation in 
power which is many orders of magnitude larger than what was 
previously calculated \cite{golwala,golubev} and 
which corresponds just to the term labeled as Eq. (\ref{pshot_1}).
Nevertheless, we shall show next that the extra terms [(\ref{pshot_2}) and 
(\ref{pshot_3})] have no effect on the observable quantities. 
The shot noise in the output power has no direct connection with 
the $NEP$ of the detector, but 
the fluctuations in the power would produce 
fluctuations in the electronic temperature, which depreciate the 
detection properties of the microbolometer. The $NEP$ is then 
defined as {\em the minimum power of the input signal that would 
produce a change in temperature equal to the square root of the 
mean square fluctuation of temperature.}

\section{Noise in the Thermal Sensing Element}

\subsection{Temperature fluctuations} \label{tf}

The detection of radiation using microbolometers is based on the 
measurement of quantities which depend on the electronic 
temperature in the detector and, eventually, on the chemical 
potential of the electrons (as is the case of NIS junctions 
used as thermometers). Therefore, the noise in the {\em total} 
electronic energy is not directly connected to the figure of 
merit of the detector, 
which is the $NEP$. 

At very low temperatures, the chemical potential of a Fermi system 
is related to the Fermi energy by the relation 
$
\mu\approx\epsilon_{\rm F}\left[ 1 - (\pi^2/12) (k_{\rm B} T / 
\epsilon_{\rm F})^2\right]
$,
where we used some obvious notations. On the other hand, the Fermi 
energy is determined by the density of particles, $n=N/V$, in 
the following way:
$
n=(2m_{\rm e} / \hbar^2)^{3/2} \epsilon_{\rm F}^{3/2}/(3\pi^2)
$. 
Other quantities of interest are the energy per particle $u$ and 
the specific heat at constant volume, 
$c_{\rm V}=(\partial u/\partial T)_{N,V={\rm constant}}$, which, 
in the low temperature limit have the expressions 
$
u =  (3/5) \epsilon_{\rm F} [1+(5\pi^2/12) 
(k_{\rm B}T / \epsilon_{\rm F})^2] 
$ and 
$
c_{\rm V} = (\pi^2/2)(k_{\rm B}^2T/\epsilon_{\rm F}) 
$.

Let us now calculate the change in the temperature of the electron 
gas when $\delta N_\epsilon$ electrons of energy 
$\epsilon + \epsilon_{\rm F}$
leave the normal metal. This loss of electrons changes 
$u$ by an amount:
\begin{equation} \label{var_u1}
\delta u = \frac{U-\delta N_\epsilon
(\epsilon+\epsilon_{\rm F})}{N-\delta N_\epsilon} - \frac{U}{N} 
\approx u \frac{\delta N_\epsilon}{N} - (\epsilon+\epsilon_{\rm F})
\frac{\delta N_\epsilon}{N}, 
\end{equation}
which can be expressed as 
\begin{equation} \label{var_u2}
\delta u = \frac{\partial u}{\partial T}\delta T + 
\frac{\partial u}{\partial \epsilon_{\rm F}}\delta \epsilon_{\rm 
F}.
\end{equation}
The change in the Fermi energy can be written as 
$\delta \epsilon_{\rm F} = -(2/3) \epsilon_{\rm F}\delta N/N$. 
Using the two equations we have for $\delta u$, [Eqs. 
(\ref{var_u1}) and (\ref{var_u2})], and neglecting the terms of 
the order $\alpha^{-2} \equiv (k_{\rm B}T/\epsilon_{\rm F})^2 
\ll 1$, we end up with an expression for the temperature variation:
\begin{equation}\label{dt_eps}
c_{\rm V}\delta T_\epsilon = -\epsilon\frac{\delta N_\epsilon}{N} .
\end{equation}
Here the subscript $\epsilon$ in the notation $\delta T_\epsilon$ 
refers to the fact that the variation in temperature is due to the 
loss of electrons from the level with energy $\epsilon + 
\epsilon_{\rm F}$. In the linear approximation, the temperature 
change is an additive quantity. Taking the time derivative on both 
sides of Eq. (\ref{dt_eps}), we can write:
\begin{equation}\label{dtt_eps}
c_{\rm V}\delta \dot{T}_\epsilon = -\epsilon\frac{\delta 
\dot{N}_\epsilon}{N}.
\end{equation}
Using the expressions for the chemical potential, the internal energy 
and Eq. (\ref{dt_eps}), it can 
be shown that the variation of $\mu$ with temperature introduces 
corrections of order $\alpha^{-2}$ to the first order terms, in all 
the following equations. Therefore in the rest of this paper we shall 
consider $\mu = \epsilon_{\rm F}$.

\subsection{Noise Equivalent Power} \label{nep}

\begin{figure}[h]
\begin{center}
\unitlength1mm\begin{picture}(60,80)(0,0)
\put(0,0){\epsfig{file=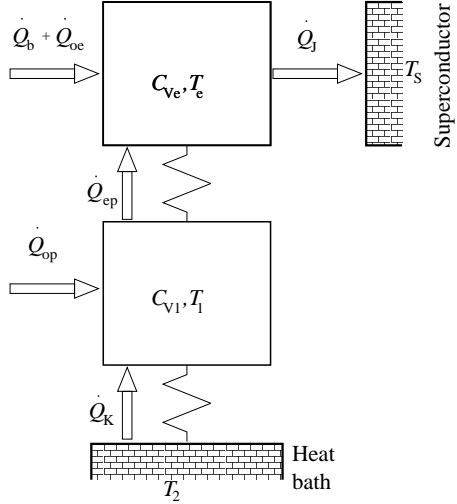,width=60mm}}
\end{picture}
\caption{A schematic drawing of the thermal sensing element. The 
two rectangular boxes represent two distinct subsystems of this 
element: the electron system and the lattice. The electron system, 
the lattice, the heat bath, and the superconductor are at the 
temperatures $T_{\rm e}$, $T_1$, $T_2$, and $T_{\rm s}$, 
respectively. In the case of direct cooling the superconductor is 
connected with the electron gas through NIS junctions, while in the 
case of indirect cooling it is used to cool down the thermal bath. 
The heat capacities of the electron gas and of the lattice are 
$C_{\rm Ve}$ and $C_{\rm V1}$, respectively. The power fluxes 
transmitted between the systems are represented in the figure by 
long arrows and are denoted by $\dot Q_{\rm J}$, 
$\dot Q_{\rm ep}$, $\dot Q_{\rm K}$. 
$\dot Q_{\rm b}$ is the bias power, while $\dot Q_{\rm oe}$ and 
$\dot Q_{\rm op}$ are the optical input powers into the electron 
system and into the lattice, respectively.}
\label{bolo}
\end{center}
\end{figure}

Having now calculated the variations in temperature 
due to the variations in the population of the energy levels, 
we can determine the total temperature fluctuation 
in the ES of the TSE. The system is represented schematically in 
Fig. \ref{bolo}. The electrons interact with the lattice, at 
temperature $T_1$, the detector lattice exchanges heat with a heat 
bath, at temperature $T_2$, and the thermal resistance between the 
lattice and the heat bath is the Kapitza resistance.

We have to explain here the notation. 
$\dot{Q}$ will be used for the power fluxes (with the directions given 
by the long arrows in Fig. \ref{bolo}) and $\dot{N}_\epsilon$ for the 
time variation of the occupation number of the electronic level of 
energy $\epsilon$ (taken with changed sign sign, as we did in the 
previous section). The subscripts are used to specify the 
processes that caused the power transfer or the time variation of the 
particle population. In this way, J refers to the particle transfer 
through the NIS junction, ep to the electron-phonon interaction 
in the TSE, K to the heat transfer through the contact between the 
lattice of the TSE and the heat bath, while oe and op refer to the 
optical input power, due to the external radiation, into the 
ES and lattice of the TSE, respectively. By $\dot{Q}_{\rm b}$ we 
denote the bias power, as shown in the figure, but this contribution 
will be disregarded until the end of the paper, where it will be 
discussed in connection with the uncertainty introduced in experiment 
by the measurement process. Moreover, 
we shall use the Greek letter $\delta$ in front of the notation 
to specify the fluctuations of a quantity and the subscript 
shot for the fluctuations of this quantity, due solely to the 
shot noise (discrete amounts transported randomly at a constant 
average rate).

If we add up all the noise terms into the power balance 
equation, disregarding the fluctuations of $T_2$ and $T_{\rm s}$, 
we arrive at the following set of equations:
\begin{eqnarray}
\left\{\begin{array}{lcl}
C_{\rm Ve}\delta\dot{T}_{\epsilon_{\em e}} &=& 
-\epsilon(\delta\dot{N}_{\epsilon_{\rm J,shot}} + 
\delta\dot{N}_{\epsilon_{\rm ep,shot}} + 
\delta\dot{N}_{\epsilon_{\rm oe,shot}}) \\ 
& & - \epsilon \frac{\partial(\dot{N}_{\epsilon_{\rm J}}
+\dot{N}_{\epsilon_{\rm ep}} + \dot{N}_{\epsilon_{\rm oe}})}{ 
\partial T_{\rm e}} \delta T_{\rm e} - \epsilon
\frac{\partial\dot{N}_{\epsilon_{\rm J}}}{\partial\mu_{\rm e}}
\delta\mu_{\rm e}- \epsilon
\frac{\partial\dot{N}_{\epsilon_{\rm ep}}}{\partial T_1}
\delta T_1 \\
C_{\rm V1}\delta\dot{T}_1 &=& -\delta \dot{Q}_{\rm ep,shot} 
+ \delta \dot{Q}_{\rm K,shot} + \delta \dot{Q}_{\rm op,shot} \\
& & -\frac{\partial (\dot{Q}_{\rm ep} - \dot{Q}_{\rm K} - 
\dot{Q}_{\rm op})}{\partial T_1}\delta T_1 - 
\frac{\partial \dot{Q}_{\rm ep}}{\partial T_{\rm e}} 
\delta T_{\rm e}
\end{array}\right. , \label{set}
\end{eqnarray}
where, transforming the summations over the energy levels into 
integrals, we use the notation $\delta T_{\rm e}\equiv\int d\epsilon \sigma_0 
\delta T_{\epsilon_{\rm e}}$ in which $\sigma_0$ is the density of the 
energy levels of the electrons in the normal metal, at Fermi energy 
(we make here the usual assumption that the energy range of interest 
is much smaller than the Fermi energy, so we can consider the density 
of states as being almost constant). Calculating the Fourier 
transformations of the two equations above and replacing 
the expression for $\delta T_1(\omega)$ obtained from the second 
equation into the first equation, we get
\begin{eqnarray}
 &i\omega C_{\rm Ve}\delta T_{\epsilon_{\rm e}}(\omega)+
\delta T_{\rm e}(\omega)\left[\epsilon
\frac{\partial(\dot{N}_{\epsilon_{\rm J}}+
\dot{N}_{\epsilon_{\rm ep}} + 
\dot{N}_{\epsilon_{\rm oe}})}{\partial T_{\rm e}}-\epsilon
\frac{\partial\dot{N}_{\epsilon_{\rm ep}}}{\partial T_1}
\frac{\partial \dot{Q}_{\rm ep}}{\partial T_{\rm e}}
\frac{1}{i\omega C_{\rm V1}+
\frac{\partial(\dot{Q}_{\rm ep} - \dot{Q}_{\rm K} - 
\dot{Q}_{\rm op})}{\partial 
T_1}}\right]& = \nonumber\\
=& -\epsilon [\delta\dot{N}_{\epsilon_{\rm J,shot}}(\omega) + 
\delta\dot{N}_{\epsilon_{\rm ep}}(\omega) + 
\delta\dot{N}_{\epsilon_{\rm oe,shot}}(\omega)] - \epsilon\frac{
\partial\dot{N}_{\epsilon_{\rm J}}}{\partial E}
\frac{\partial\epsilon_{\rm F}}{\partial N}
\frac{\delta\dot{N}_{\rm J}(\omega)}{i\omega} & 
\nonumber\\
& -\epsilon\frac{\partial\dot{N}_{\epsilon_{\rm ep}}}{\partial T_1}
\frac{-\delta \dot{Q}_{\rm ep,shot}(\omega) + 
\delta \dot{Q}_{\rm K,shot}(\omega) + 
\delta \dot{Q}_{\rm op,shot}(\omega)}{i\omega C_{\rm V1} 
+ \frac{\partial(\dot{Q}_{\rm ep} - \dot{Q}_{\rm K} 
- \dot{Q}_{\rm op})}{\partial T_1}} \, .
&  \label{dt_omega}
\end{eqnarray}
We introduce the notation:
\begin{eqnarray}
\Delta_{T_\epsilon}(\omega)&\equiv& 
i\omega C_{\rm Ve}\delta T_{\epsilon_{\rm e}}(\omega) 
+ \delta T_{\rm e}(\omega) \left[\epsilon
\frac{\partial (\dot{N}_{\epsilon_{\rm J}} +
\dot{N}_{\epsilon_{\rm ep}} + \delta 
\dot{N}_{\epsilon_{\rm oe}})}{\partial T_{\rm e}} \right. 
\nonumber\\
& & \left. -\epsilon \frac{\partial\dot{N}_{\epsilon_{\rm ep}}}{ 
\partial T_1} \frac{\partial \dot{Q}_{\rm ep}}{ 
\partial T_{\rm e}} \frac{1}{i\omega C_{\rm V1}+
\frac{\partial (\dot{Q}_{\rm ep} - \dot{Q}_{\rm K} 
- \dot{Q}_{\rm op})}{\partial T_1}}\right] . \label{delta_e} 
\end{eqnarray}
We can now integrate both sides of Eq. (\ref{delta_e}) and write 
\begin{eqnarray}
\Delta_T(\omega) &=& \delta T_{\rm e}(\omega) \left[ 
i\omega C_{\rm Ve} + \frac{\partial}{\partial T_{\rm e}} 
(\dot{Q}_{\rm J} - \dot{Q}_{\rm ep} - \dot{Q}_{\rm oe})\right. 
\nonumber \\
& &\left.+\frac{1}{i\omega C_{\rm V1}+\frac{\partial 
(\dot{Q}_{\rm ep} - \dot{Q}_{\rm K} - \dot{Q}_{\rm op})}{
\partial T_1}}\frac{\partial \dot{Q}_{\rm ep}}{\partial T_{\rm e}}
\frac{\partial \dot{Q}_{\rm ep}}{\partial T_1}\right] . 
\label{delta1}
\end{eqnarray}
Integrating the r. h. s. of Eq. (\ref{dt_omega}) we obtain 
another expression for $\Delta_T(\omega)$:
\begin{eqnarray}
\Delta_T(\omega) &=& -\delta \dot{Q}_{\rm J,shot}(\omega) + 
\delta \dot{Q}_{\rm ep,shot}(\omega) + 
\delta \dot{Q}_{\rm oe,shot}(\omega) \nonumber \\ 
& & - \frac{\partial\epsilon_{\rm F}}{\partial N}
\frac{\partial \dot{Q}_{\rm J}}{\partial E}\frac{\delta
\dot{N}_{\rm J,shot}(\omega)}{i\omega} \nonumber \\
& &+\frac{\partial \dot{Q}_{\rm ep}}{\partial T_1} 
\frac{-\delta \dot{Q}_{\rm ep,shot}(\omega) + 
\delta \dot{Q}_{\rm K,shot}(\omega) + 
\delta \dot{Q}_{\rm op,shot}(\omega)}{i\omega C_{\rm V1}+
\frac{\partial(\dot{Q}_{\rm ep} - \dot{Q}_{\rm K} - 
\dot{Q}_{\rm op})}{\partial T_1}} \, .
\label{delta2}
\end{eqnarray}
Using Eqs. (\ref{delta1}) and (\ref{delta2}) we obtain the 
following expressions for the mean square value of $\Delta_T(\omega)$: 
\begin{eqnarray}
\langle\Delta_T^2(\omega)\rangle &=& 
\langle\delta^2 T_{\rm e}\rangle_\omega \left|i\omega C_{\rm Ve}+
\frac{\partial}{\partial T_{\rm e}} (\dot{Q}_{\rm J} - 
\dot{Q}_{\rm ep} - \dot{Q}_{\rm oe})\right. \nonumber \\
& &\left.+\frac{1}{i\omega C_{\rm V1}+\frac{\partial
(\dot{Q}_{\rm ep} - \dot{Q}_{\rm K} - \dot{Q}_{\rm op})}{
\partial T_1}}\frac{\partial \dot{Q}_{\rm ep}}{\partial T_{\rm e}}
\frac{\partial \dot{Q}_{\rm ep}}{\partial T_1}\right|^2 
\label{delta1p}\\
&=&\langle\delta^2 \dot{Q}_{\rm ep,shot}\rangle_\omega\left|1+
\frac{\partial \dot{Q}_{\rm ep}}{\partial T_1}
\frac{1}{i\omega C_{\rm V1}+\frac{\partial 
(\dot{Q}_{\rm ep} - \dot{Q}_{\rm K} - \dot{Q}_{\rm op})}{\partial 
T_1}}\right|^2 \nonumber \\
& & + \left[ \langle\delta^2 \dot{Q}_{\rm K,shot}\rangle
_\omega
+ \langle\delta^2 \dot{Q}_{\rm op,shot}\rangle_\omega \right]
\left(\frac{\partial \dot{Q}_{\rm ep}}{\partial T_1}\right)^2 
\nonumber \\
 & & \times \left\{ \omega^2C_{\rm V1}^2+\left[\frac{\partial 
(\dot{Q}_{\rm ep} - \dot{Q}_{\rm K} - \dot{Q}_{\rm op})}{
\partial T_1}\right]^2 \right\}^{-1} \nonumber\\
& & + \langle \delta^2 \dot{Q}_{\rm oe,shot}\rangle_\omega 
\nonumber \\
& & + \left\langle \left| \left( \delta \dot{Q}_{\rm J,shot}+
\frac{\partial\epsilon_{\rm F}}{\partial N}\frac{\partial 
\dot{Q}_{\rm J}}{\partial E} \frac{\delta \dot{N}_{\rm 
J,shot}}{i\omega} \right) (\omega) \right|^2 \right\rangle \, .
\label{delta2p}
\end{eqnarray}
The last term of Eq. (\ref{delta2p}), say $\Upsilon(\omega)$, can 
be calculated. The correlation function between 
$\delta \dot{Q}_{\rm J,shot} (\omega)$ and 
$\delta \dot{N}_{\rm J,shot} (\omega) / i\omega$ 
gives no contribution due to the $\pi/2$ phase difference, 
and the final result is:
\begin{equation}\label{upsi}
\Upsilon(\omega)=\langle\delta^2\dot{Q}_{\rm 
J,shot}\rangle_\omega+
\frac{1}{\omega^2}\left(\frac{\partial\epsilon_{\rm F}}{\partial N}
\frac{\partial \dot{Q}_{\rm J}}{\partial E}\right)^2
\langle\delta^2\dot{N}_{\rm J,shot}\rangle_\omega
\end{equation}

From the expressions (\ref{delta1p}) and (\ref{delta2p}), using the 
result (\ref{upsi}), one can calculate the spectral density of the 
temperature noise.

To determine the amplitude of the signal power that has to 
be introduced into the detector, to produce changes in the temperature 
of the ES equal to the fluctuations calculated above, we 
write a set of equations similar to (\ref{set}):
\begin{eqnarray}
C_{\rm Ve}\delta\dot{T}_{\rm e} &=& \dot{Q}_{\rm s} - 
\dot{Q}_{\rm J0} + \dot{Q}_{\rm ep0} + \dot{Q}_{\rm oe0} 
\label{rad1} \\
 & & - \frac{\partial (\dot{Q}_{\rm J} - \dot{Q}_{\rm ep} - 
\dot{Q}_{\rm oe})}{\partial T_{\rm e}}
\delta T_{\rm e} + \frac{\partial \dot{Q}_{\rm ep}}{\partial 
T_1}\delta T_1 \, , \nonumber \\
C_{\rm V1}\delta\dot{T}_1 &=& \dot{Q}_{\rm K0} + 
\dot{Q}_{\rm op0} - \dot{Q}_{\rm ep0}+
\frac{\partial(\dot{Q}_{\rm K} + \dot{Q}_{\rm op0} - 
\dot{Q}_{\rm ep})}{\partial T_1}\delta T_1 \nonumber \\
 & & - \frac{\partial \dot{Q}_{\rm ep}}{\partial T_{\rm e}} 
\delta T_{\rm e} \, .
\label{rad2}
\end{eqnarray}
Here $\dot{Q}_{\rm s}$ is the power of the input signal into 
the TSE. The subscript 0 is added to the usual subscripts 
to denote the equilibrium values of the powers defined in Fig. 
\ref{bolo}. Therefore, $-\dot{Q}_{\rm J0} + \dot{Q}_{\rm ep0} + 
\dot{Q}_{\rm oe0} = \dot{Q}_{\rm K0} + \dot{Q}_{\rm op0} - 
\dot{Q}_{\rm ep0}=0$. Calculating the Fourier transformations 
of Eqs. (\ref{rad1}) and (\ref{rad2}) 
and replacing $\delta T_1(\omega)$ from Eq. (\ref{rad2}) into Eq. 
(\ref{rad1}), we obtain the {\em Noise Equivalent Power}, 
as defined in the end of Section \ref{intro}:
\begin{equation}
NEP^2 \equiv |\dot{Q}_{\rm s}(\omega)|^2 = |\Delta_T^2(\omega)| 
\label{nis}.
\end{equation}
Therefore, $NEP$ can be calculated directly from Eq. 
(\ref{delta2p}), with the use of Eq. (\ref{upsi}). In this 
way we observe that when the detector is cooled 
indirectly ($\dot{Q}_{\rm J}\equiv 0$ and $T_2\approx T_1 
\approx T_{\rm e}$), $\Upsilon \equiv 0$ and $NEP$ is 
lower. We shall evaluate the total $NEP$ in the next 
sections. 

\subsection{Calculation of the shot noise terms} \label{n_terms}

\paragraph{Electron-phonon shot noise.}

The heat power transferred between the ES and the lattice 
of the TSE, $\dot{Q}_{\rm ep}$, due to the electron-phonon 
interaction, can be calculated using the formula 
\begin{equation}\label{wellst}
\dot{Q}_{\rm ep} = \Sigma_{\rm ep} \Omega (T_1^5-T_{\rm e}^5) ,
\end{equation}
where $\Omega$ is the volume of the TSE and $\Sigma_{\rm ep}$ is the 
electron-phonon coupling constant \cite{wellstood}. If, for example, 
the TSE is made of copper, then 
$\Sigma_{\rm ep} \approx 4\ {\rm nW/K^5\ \mu m^3}$ 
\cite{jukka}. Using the same model as in Ref. \cite{wellstood}, the 
shot noise of $\dot{Q}_{\rm ep}$ has been evaluated in Ref. 
\cite{golwala}. An approximative expression for this power shot noise, 
namely 
\begin{equation}\label{gol}
\langle\delta^2\dot{Q}_{\rm ep,shot}\rangle_\omega \approx
5\Sigma_{\rm ep} \Omega (T_{\rm e}^6+T_1^6) ,
\end{equation}
has been used in Ref. \cite{golubev} (although a factor of two 
was introduced by mistake there). The formula above is 
identical to the exact expression for $T_{\rm e} = T_1$ and 
deviates from it with less than 2\% for any $T_{\rm e} \le T_1$. 
Therefore we shall use Eq. (\ref{gol}) in the rest of this paper, 
since its accuracy is good enough for our purposes.

\paragraph{Kapitza shot noise.}

We assume that the dynamics of the detector lattice is well 
described by a three-dimensional distribution of acoustic phonons, 
with sound velocity $v$ (the same for all three phonon modes). 
If the heat bath consists of a dielectric membrane, with the sound 
velocity $v_{\rm m}$, then the phonon flux that penetrates through 
the separation surface, of area $S$, from the detector into the membrane 
has the expression:
\begin{eqnarray} \begin{array}{lcl}
\dot{N}_{\rm K} &=& 
\left\{ \begin{array}{l}
S \frac{3v}{4\pi^2}\left(\frac{v}{v_{\rm m}}\right)^2
\left(\frac{k_{\rm B}T}{\hbar v}\right)^3 \zeta(3)  t,\ {\rm 
for}\ v\le v_{\rm m} \, , \\ \\
S \frac{3v}{4\pi^2} \left( \frac{k_{\rm B}T}{\hbar v}\right)^3 
\zeta(3)  t,\ {\rm for}\ 
v > v_{\rm m} \, , 
\end{array} \right.
\end{array}
\label{fononi}
\end{eqnarray}
where $t$ is the transmission coefficient, related to the acoustic 
impedances of the detector lattice, $Z$, and of the membrane, 
$Z_{\rm m}$, by the relation $t=4ZZ_{\rm m}(Z+Z_{\rm m})^{-2}$ 
\cite{pobel}. The energy flux from the heat bath into the detector is
\begin{equation}\label{energy}
\dot{Q}_{\rm K} = S\frac{9\zeta(4)}{4\pi^2} t 
\frac{k_{\rm B}^4}{v_{\rm m}^2\hbar^3}(T_2^4-T_1^4) 
\equiv \Sigma_{\rm K}S(T_2^4-T_1^4) ,
\end{equation}
and the power shot noise due to the quantisation of the phonon 
energy is 
\begin{equation}\label{shot}
\langle\delta^2 \dot{Q}_{\rm K,shot}\rangle_\omega = 
8\frac{\zeta(5)}{\zeta(4)} \Sigma_{\rm K} S (T_2^5 + T_1^5),
\end{equation}
which, when $T_1=T_2$, assumes the form 
$
\langle\delta^2 \dot{Q}_{\rm K,shot}\rangle_\omega = 
(\zeta(5)/\zeta(4)) 4k_{\rm B}T^2G_{\rm K}\approx 
4k_{\rm B}T^2G_{\rm K},
$
where $G_{\rm K}\equiv |\partial \dot{Q}_{\rm K}/ \partial 
T_1|_{T_1=T_2}$ is the Kapitza conductance. 

\paragraph{Optical input power shot noise.} 

Let us suppose that the detector admits a frequency band 
$\Delta f_{\rm o}$ from the 
electromagnetic spectrum within which the quantum efficiency is unity.
If we denote by $n (\omega_{\rm o})$ the 
density of phonons with the angular frequency $\omega_{\rm o}$ 
in a spatial region around the detector, then the 
energy flux and the number flux of photons on the detector are 
$\dot{Q}_{\rm o}(\omega_{\rm o}) = S\hbar \omega_{\rm o} c n(\omega_{\rm o}) 
\Delta \omega_{\rm o}/4$ and $\dot{N}_{\rm o}(\omega_{\rm o}) = 
S c n(\omega_{\rm o}) \Delta \omega_{\rm o}/4$, respectively, where 
$\omega_{\rm o} \equiv 2\pi f_{\rm o}$, as usual, $c$ is the 
velocity of light, and 
we assume that $\Delta \omega_{\rm o} \ll \omega_{\rm o}$. The electromagnetic 
radiation interacts with the ES, therefore we shall take 
$\dot{Q}_{\rm o}(\omega_{\rm o}) = \dot{Q}_{\rm oe}(\omega_{\rm o})$ 
and $\dot{Q}_{\rm op}(\omega_{\rm o})=0$. Since 
$n(\omega_{\rm o}) = (1/\pi^2c^3)\omega_{\rm o}^2 
(\exp{(\hbar \omega_{\rm o} /k_{\rm B}T)} -1)^{-1}$ (see for example 
Chap. 10 in Ref. \cite{mandl}), where $T$ is 
the temperature of the background radiation, and if we assume that 
each photon produces a $\delta$ peak of power into the ES,
we obtain the following expressions for the input power and the 
shot noise into the ES of the TSE, respectively:
\begin{equation} \label{q_oe} 
\dot{Q}_{\rm oe}(\omega_{\rm o}) = \frac{S}{4}\frac{\hbar}{\pi^2 c^2} 
\frac{\omega_{\rm o}^3 \, \Delta\omega_{\rm o}}{
\exp{(\hbar \omega_{\rm o} /k_{\rm B}T)} -1} 
\end{equation}
and
\begin{equation} \label{n_oe} 
\langle\delta^2 \dot{Q}_{\rm oe,shot}(\omega_{\rm o})\rangle_\omega = 
\frac{S}{2}\left(\frac{\hbar}{\pi c} \right)^2 
\frac{\omega_{\rm o}^4 \, \Delta\omega_{\rm o}}{
\exp{(\hbar \omega_{\rm o} /k_{\rm B}T)} -1} \, .
\end{equation}
The radiation that influences the astrophysical measurements 
is the cosmic background radiation, which corresponds to 
$T \equiv T_{\rm b} \approx$ 3 K.

\paragraph{Shot noise in the junction.} 

As anticipated in 
Section \ref{intro}, the shot noise fluctuation of the total power 
extracted through the NIS junction has no relevance to our problem. The 
calculations in Sections \ref{tf} and \ref{nep} showed in a rigorous 
manner that the quantity of interest is $\dot{Q}_{\rm J}$, given by 
Eq. (\ref{power_c}). The shot noise fluctuation of this 
quantity, which enters directly into the calculation of $NEP$, 
has the expression 
\begin{eqnarray}
\langle\delta^2 \dot{Q}_{\rm J,shot}\rangle_\omega 
&=& \frac{2}{e^2R_{\rm T}}\int_\Delta^\infty d\epsilon
g(\epsilon)\bigg\{(\epsilon-eV)^2 \nonumber \\
& &\times [f(\epsilon-eV, T_{\rm e})(1-2f(\epsilon, T_{\rm s}))+
f(\epsilon, T_{\rm s})] \nonumber \\
& & + (\epsilon+eV)^2[f(\epsilon+eV, T_{\rm e})(1-2f(\epsilon, T_{\rm s}))+
f(\epsilon, T_{\rm s})]\bigg\} \, , \label{pshot_f} 
\end{eqnarray}
which is the term labeled by (\ref{pshot_1}) in Section \ref{intro}.
The fluctuation of the particle flux, 
$\langle\delta^2\dot{N}_{\rm J,shot}\rangle_\omega$, 
that enters 
the expression of $\Upsilon(\omega)$, can be calculated in a similar 
way as $\langle\delta^2 \dot{Q}_{\rm J,shot}\rangle_\omega$, 
with the result 
\begin{eqnarray}
\langle\delta^2 \dot{N}_{\rm J,shot}\rangle_\omega 
&=& \frac{2}{e^2R_{\rm T}}\int_\Delta^\infty d\epsilon
g(\epsilon)\bigg\{ [f(\epsilon-eV, T_{\rm e}) + 
f(\epsilon+eV, T_{\rm e})] \nonumber \\
& & \times [1-2f(\epsilon, T_{\rm s})] + 2f(\epsilon, T_{\rm s})] 
\bigg\} \, . \label{nshot_f} 
\end{eqnarray}

We have written down explicitly all the equations needed for the 
calculation of $NEP$. We are now left with the task of calculating 
the parameters of the refrigerating junctions for specific working 
conditions. Since the formulae involved are rather complicated 
\cite{trap,teza} and 
the microbolometers work at low temperatures, we give below some 
useful analytical approximations, valid for this range of temperatures. 

\section{Approximative analytical expressions for the calculation 
of the junctions parameters} \label{analytic}

In order to calculate the efficiency of the microbolometer 
we have to know the parameters of the refrigerating junctions 
for given working conditions. The physical phenomena that 
take place in NIS microrefrigerators have been presented in a 
series of publications (see for example Refs. 
\cite{nahum3,jukka,jochum,trap,averin,teza}). Yet, the formulae 
involved are rather complicated and, in the general case, have to 
be calculated numerically. The metal that had been used extensively as 
a superconductor in the construction of the NIS refrigerating 
junctions is Al, which has an energy gap $\Delta \approx 200 \ 
{\rm \mu eV}$ at temperatures much lower than the critical 
one. Fortunately, the temperature range 
of interest for us ($T \le 0.3$ K) is well below the critical temperature 
and since in this range $\Delta/k_{\rm B}T \gg 1$,  we shall give here
a general method to calculate analytical approximations for all the 
formulae presented in connection with the NIS junctions.

\subsection{General formulae} \label{general}

In the previous sections we saw that all the formulae needed for the 
quantitative evaluation of the processes in NIS junctions 
[see Eqs. (\ref{charge_c}), (\ref{excitation_c}), 
(\ref{power_c}), (\ref{pshot_1}), (\ref{pshot_2}), 
(\ref{pshot_3}), (\ref{pshot_f}), and (\ref{nshot_f})] have the 
general form:
\begin{eqnarray}
^{(r)}F^{(s)}_l &\equiv& \frac{1+\theta (s)}{e^2R_{\rm T}} 
\int_\Delta^\infty d\epsilon\, 
\frac{\epsilon}{\sqrt{\epsilon^2-\Delta^2}} \bigg\{ (\epsilon-E)^l
[f(\epsilon-E,T_{\rm e}) (1-f(\epsilon,T_{\rm s})) \nonumber\\
& & + s (1-f(\epsilon-E,T_{\rm e}))f(\epsilon,T_{\rm s})] + r  
(\epsilon+E)^l \label{fs1} \\
& &\times[f(\epsilon+E,T_{\rm e}) (1-f(\epsilon,T_{\rm s}))
+ s (1-f(\epsilon+E,T_{\rm e}))f(\epsilon,T_{\rm s})]\bigg\} 
\nonumber\\
&=& \frac{1+\theta (s)}{e^2R_{\rm T}}\int_\Delta^\infty d\epsilon\, 
\frac{\epsilon}{\sqrt{\epsilon^2-\Delta^2}} \label{fs2} \\
& &\times \bigg\{ (\epsilon-E)^l [f(\epsilon-E,T_{\rm e}) + s 
f(\epsilon,T_{\rm s}) - (1 + s)f(\epsilon-E,T_{\rm e}) 
f(\epsilon,T_{\rm s})] \nonumber\\
& & + r (\epsilon+E)^l [f(\epsilon+E,T_{\rm e}) + s  
f(\epsilon,T_{\rm s}) - (1 + s)f(\epsilon+E,T_{\rm e}) 
f(\epsilon,T_{\rm s})]\bigg\} \, .
\nonumber
\end{eqnarray}
The parameter $l$ takes one of the values 0, 1, or 2, while 
$r$ and $s$ can be $+1$ or $-1$. The function $\theta (s)$ is 
the Heaviside step function.
For example $^{(+)}F^{(+)}_2 \equiv \langle \delta^2 
\dot{Q}_{\rm J, shot}\rangle_\omega$, 
$^{(+)}F^{(+)}_0\equiv \langle \delta^2 \dot{N}_{\rm J,shot} 
 \rangle_\omega$, and $^{(+)}F^{(-)}_1\equiv \dot{Q}_{\rm J}$.
The integral in Eq. (\ref{fs2}) splits up in an obvious way into 
six terms. We shall evaluate them one by one in 
the approximation 
\begin{equation}\label{aprox_def}
A_{\rm e}\equiv\frac{\Delta}{k_{\rm B}T_{\rm e}} \ge 
A_{\rm s}\equiv\frac{\Delta}{k_{\rm B}T_{\rm s}}\gg 1 .
\end{equation}
Other notations that we shall use are: 
$\beta \equiv 1/k_{\rm B}T$, where any subscripts attached to $T$ 
will be transferred to $\beta$, 
$\delta \equiv \Delta-E$, 
$a_{\rm e} \equiv \beta_{\rm e}\delta$,  
$a_{\rm s} \equiv \beta_{\rm s}\delta $, 
$B_{\rm e} \equiv \beta_{\rm e}(\Delta+E)\gg 1$, and 
$B_{\rm s} \equiv \beta_{\rm s}(\Delta+E)\gg 1$.

The analytical approximations are possible due to the exponential 
dependence of the Fermi functions on energy, and are based on the 
Taylor expansion of the function $1/\sqrt{1+x}$, around $x=0$. 
If we also write $(a+b)^l = a^l + 
la^{l-1}b + [l(l-1)/2] a^{l-2}b^2$ (which is exact for $l=0,1,2$), 
we can calculate general analytical expressions for the 
six integrals in Eq. (\ref{fs2}). In what follows we shall be interested 
just in the two highest order terms in $A_{\rm e}$, $A_{\rm s}$, 
$B_{\rm e}$, and $B_{\rm s}$ in each of the integrals mentioned. 
For each $l$=0, 1, or 2 these highest order terms can be extracted 
rigorously from the general formulae given below, 
but in some cases the remaining lower order terms do not have 
the complete form (therefore we advise the reader to recalculate these 
lower orders, if they are necessary).

Let us now start to calculate each of the integrals from Eq. (\ref{fs2}). 
In these calculations, the values of the parameters $r$ and $s$ 
have no relevance, so they will be omitted. They should be 
introduced just in the end, 
when the expression of $^{(r)}F^{(s)}_l$ is to be reconstructed.
The first term is 
\begin{eqnarray}
I_{1_l}&\equiv& \int_\Delta^\infty d\epsilon\, 
\frac{\epsilon}{\sqrt{\epsilon^2-\Delta^2}} 
\frac{(\epsilon-E)^l}{\exp{[\beta_{\rm e} (\epsilon-E)]} + 1} 
\nonumber \\
&=& \frac{(k_{\rm B}T_{\rm e})^{l+1}}{\sqrt{2A_{\rm e}}} 
\left\{A_{\rm e}\left[\Gamma(l+1/2)g_{l+1/2}(a_{\rm e}) \right. 
\right. 
\nonumber\\
& &\left.\left. + l\Gamma(l-1/2)a_{\rm e}g_{l-1/2}(a_{\rm e}) + 
\frac{l(l-1)}{2}\Gamma(l-3/2)a^2_{\rm e}g_{l-3/2}(a_{\rm 
e})\right] 
\right. \nonumber \\
& &\left. + \frac{3}{4}\left[ \Gamma(l+3/2) g_{l+3/2}(a_{\rm e}) 
+ 
l \Gamma(l+1/2) a_{\rm e} g_{l+1/2}(a_{\rm e}) 
\right. \right. \nonumber \\
& &\left. \left. + \frac{l(l-1)}{2} \Gamma(l-1/2) a^2_{\rm e} 
g_{l-1/2}(a_{\rm e}) \right] \right\} , \nonumber
\end{eqnarray}
where the functions $g_l(\alpha)$ are the 
$l^{\rm th}$ order polylogarithmic functions of argument 
$-e^{-\alpha}$, $g_l(\alpha) \equiv [1/\Gamma(l)] 
\int_0^\infty t^{l-1}(e^{t+\alpha}+1)^{-1}\, dt$ 
(see, for example, Refs. \cite{lee1} and references therein for more 
details). 

In the second term:
\begin{eqnarray}
I_{2_l}&\equiv& \int_\Delta^\infty d\epsilon\, 
\frac{\epsilon}{\sqrt{\epsilon^2-\Delta^2}} 
\frac{(\epsilon-E)^l}{\exp{(\beta_{\rm s} \epsilon)} + 1} \nonumber 
\\
&=& \frac{(k_{\rm B}T_{\rm s})^{l+1}}{\sqrt{2A_{\rm s}}} 
\left\{A_{\rm s}\left[\Gamma(l+1/2)g_{l+1/2}(A_{\rm s}) \right. 
\right. 
\nonumber\\
& &\left.\left. + l\Gamma(l-1/2)a_{\rm s}g_{l-1/2}(A_{\rm s}) + 
\frac{l(l-1)}{2}\Gamma(l-3/2)a^2_{\rm s}g_{l-3/2}(A_{\rm 
s})\right] 
\right. \nonumber \\
& &\left. + \frac{3}{4}\left[ \Gamma(l+3/2) g_{l+3/2}(A_{\rm s}) 
+ 
l \Gamma(l+1/2) a_{\rm s} g_{l+1/2}(A_{\rm s}) 
\right. \right. \nonumber \\
& &\left. \left. + \frac{l(l-1)}{2} \Gamma(l-1/2) a^2_{\rm s} 
g_{l-1/2}(A_{\rm s}) \right] \right\} , \nonumber
\end{eqnarray}
since $g_l(\alpha)=e^{-\alpha}\left[1+{\mathcal O}
\left(e^{-\alpha}\right)\right]$, for any $l$, if $\alpha\gg 1$, 
we shall use the approximation $g_l(A_{\rm s})=e^{-A_{\rm s}}$, 
so: 
\begin{eqnarray}
I_{2_l}&=& \frac{(k_{\rm B}T_{\rm s})^{l+1}}{\sqrt{2A_{\rm s}}} 
e^{-A_{\rm s}}
\left\{A_{\rm s}\left[\Gamma(l+1/2) \right. \right. 
\nonumber\\
& &\left.\left. + l\Gamma(l-1/2)a_{\rm s} + 
\frac{l(l-1)}{2}\Gamma(l-3/2)a^2_{\rm s}\right] 
\right. \nonumber \\
& &\left. + \frac{3}{4}\left[ \Gamma(l+3/2) + 
l \Gamma(l+1/2) a_{\rm s} 
+ \frac{l(l-1)}{2} \Gamma(l-1/2) a^2_{\rm s} 
\right] \right\} \, . \nonumber
\end{eqnarray}

Within the same approximation, the third term, 
\[
I_{3_l} \equiv 2\int_\Delta^\infty d\epsilon\, 
\frac{\epsilon}{\sqrt{\epsilon^2-\Delta^2}} 
\frac{(\epsilon-E)^l}{\exp{[\beta_{\rm e} (\epsilon-E)]} + 1}
\frac{1}{\exp{(\beta_{\rm s} \epsilon)} + 1} \, , 
\]
can be written in the form:
\begin{eqnarray}
I_{3_l}&=& 2e^{-A_{\rm s}}
\frac{(k_{\rm B}T_{\rm e})^{l+1}}{\sqrt{2A_{\rm e}}} 
\sum_{k=0}^\infty \frac{(-1)^k}{k!} \left(\frac{T_{\rm e}}{T_{\rm 
s}} 
\right)^k
\left\{A_{\rm e}\left[\Gamma(l+k+1/2)g_{l+k+1/2}(a_{\rm e}) 
\right. \right. 
\nonumber\\
& &\left.\left. + l\Gamma(l+k-1/2)a_{\rm e}g_{l+k-1/2}(a_{\rm e}) 
+ 
\frac{l(l-1)}{2}\Gamma(l+k-3/2)a^2_{\rm e}g_{l+k-3/2}(a_{\rm 
e})\right] 
\right. \nonumber \\
& &\left. + \frac{3}{4}\left[ \Gamma(l+k+3/2) g_{l+k+3/2}(a_{\rm 
e}) + 
l \Gamma(l+k+1/2) a_{\rm e} g_{l+k+1/2}(a_{\rm e}) 
\right. \right. \nonumber \\
& &\left. \left. + \frac{l(l-1)}{2} \Gamma(l+k-1/2) a^2_{\rm e} 
g_{l+k-1/2}(a_{\rm e}) \right] \right\} . \label{I3}
\end{eqnarray}
If $T_{\rm s}\gg T_{\rm e}$ we may just take the first term in the 
summation 
and obtain
\begin{equation}\label{i3_approx}
I_{3_l}\approx 2e^{-\beta_{\rm s}\Delta}I_{1_l} \, .
\end{equation}
This is a reasonable approximation, since for a typical case of 
$T_{\rm e}\approx 0.1$ K and $T_{\rm s}\approx 0.4\ {\rm K} > 
T_2$ \cite{teza} 
$A_{\rm s} \approx 6$ and $T_{\rm s}/T_{\rm e} \approx 4$. In such 
a case $I_{3_l} \ll I_{1_l}$ so there is no point in taking higher 
order terms in Eq. (\ref{i3_approx}). 
If $T_{\rm s} = T_{\rm e}$, the infinite summations in Eq. (\ref{I3}) 
do not converge. In such a case we can make use of the general 
inequality $I_{3_l} < 2e^{-\beta_{\rm s}\Delta}I_{1_l}$, valid for 
any $T_{\rm s}$, and eventually neglect this term, since in the cases 
of practical interest $A_{\rm e} > 20$.

The fourth term is
\begin{eqnarray}
I_{4_l}&\equiv& \int_\Delta^\infty d\epsilon\, 
\frac{\epsilon}{\sqrt{\epsilon^2-\Delta^2}} 
\frac{(\epsilon+E)^l}{\exp{[\beta_{\rm e} (\epsilon+E)]} + 1} 
\nonumber \\
&=& \frac{(k_{\rm B}T_{\rm e})^{l+1}}{\sqrt{2A_{\rm e}}} 
\left\{A_{\rm e}\left[\Gamma(l+1/2)g_{l+1/2}(B_{\rm e}) \right. 
\right. 
\nonumber\\
& &\left.\left. + l\Gamma(l-1/2)B_{\rm e}g_{l-1/2}(B_{\rm e}) + 
\frac{l(l-1)}{2}\Gamma(l-3/2)B^2_{\rm e}g_{l-3/2}(B_{\rm 
e})\right] 
\right. \nonumber \\
& &\left. + \frac{3}{4}\left[ \Gamma(l+3/2) g_{l+3/2}(B_{\rm e}) 
+ 
l \Gamma(l+1/2) B_{\rm e} g_{l+1/2}(B_{\rm e}) 
\right. \right. \nonumber \\
& &\left. \left. + \frac{l(l-1)}{2} \Gamma(l-1/2) B^2_{\rm e} 
g_{l-1/2}(B_{\rm e}) \right] \right\} , \nonumber \\
&\approx& e^{-B_{\rm e}} 
\frac{(k_{\rm B}T_{\rm e})^{l+1}}{\sqrt{2A_{\rm e}}} \nonumber \\
& & \times \left\{A_{\rm e} \left[\Gamma(l+1/2) + 
l\Gamma(l-1/2)B_{\rm e} + 
\frac{l(l-1)}{2}\Gamma(l-3/2)B^2_{\rm e} \right] 
\right. \nonumber \\
& &\left. + \frac{3}{4}\left[ \Gamma(l+3/2) + 
l \Gamma(l+1/2) B_{\rm e} + \frac{l(l-1)}{2} \Gamma(l-1/2) B^2_{\rm 
e}) 
\right] \right\} . \nonumber
\end{eqnarray}

The fifth term is 
\begin{eqnarray}
I_{5_l}&\equiv& \int_\Delta^\infty d\epsilon\, 
\frac{\epsilon}{\sqrt{\epsilon^2-\Delta^2}} 
\frac{(\epsilon+E)^l}{\exp{(\beta_{\rm s} \epsilon)} + 1} \nonumber 
\\
&=& \frac{(k_{\rm B}T_{\rm s})^{l+1}}{\sqrt{2A_{\rm s}}} 
\left\{A_{\rm s}\left[\Gamma(l+1/2)g_{l+1/2}(A_{\rm s}) \right. 
\right. 
\nonumber\\
& &\left.\left. + l\Gamma(l-1/2)B_{\rm s}g_{l-1/2}(A_{\rm s}) + 
\frac{l(l-1)}{2}\Gamma(l-3/2)B^2_{\rm s}g_{l-3/2}(A_{\rm 
s})\right] 
\right. \nonumber \\
& &\left. + \frac{3}{4}\left[ \Gamma(l+3/2) g_{l+3/2}(A_{\rm s}) 
+ 
l \Gamma(l+1/2) B_{\rm s} g_{l+1/2}(A_{\rm s}) 
\right. \right. \nonumber \\
& &\left. \left. + \frac{l(l-1)}{2} \Gamma(l-1/2) B^2_{\rm s} 
g_{l-1/2}(A_{\rm s}) \right] \right\}  \nonumber\\
&\approx& \frac{(k_{\rm B}T_{\rm s})^{l+1}}{\sqrt{2A_{\rm s}}} 
e^{-A_{\rm s}}
\left\{A_{\rm s}\left[\Gamma(l+1/2) \right. \right. 
\nonumber\\
& &\left.\left. + l\Gamma(l-1/2)B_{\rm s} + 
\frac{l(l-1)}{2}\Gamma(l-3/2)B^2_{\rm s}\right] 
\right. \nonumber \\
& &\left. + \frac{3}{4}\left[ \Gamma(l+3/2) + 
l \Gamma(l+1/2) B_{\rm s} 
+ \frac{l(l-1)}{2} \Gamma(l-1/2) B^2_{\rm s} 
\right] \right\}  \nonumber
\end{eqnarray}
and the last term is
\begin{eqnarray}
I_{6_l} &\equiv& 2\int_\Delta^\infty d\epsilon\, 
\frac{\epsilon}{\sqrt{\epsilon^2-\Delta^2}} 
\frac{(\epsilon+E)^l}{\exp{[\beta_{\rm e} (\epsilon+E)]} + 1}
\frac{1}{\exp{(\beta_{\rm s} \epsilon)} + 1} \nonumber \\
 &\approx& 2e^{-(A_{\rm s}+B_{\rm e})}
\frac{(k_{\rm B}T_{\rm e})^{l+1}}{\sqrt{2A_{\rm e}}} 
\left\{A_{\rm e}\left[ 
\frac{\Gamma(l+1/2)}{(1+T_{\rm e}/T_{\rm s})^{l+1/2}} + 
l\frac{\Gamma(l-1/2)}{(1+T_{\rm e}/T_{\rm s})^{l-1/2}} 
B_{\rm e} \right.\right. \nonumber \\
& & \left. \left. + \frac{l(l-1)}{2}
\frac{\Gamma(l-3/2)}{(1+T_{\rm e}/T_{\rm s})^{l-1/2}} 
B^2_{\rm e} \right] + \frac{3}{4}\left[ 
\frac{\Gamma(l+3/2)}{(1+T_{\rm e}/T_{\rm s})^{l+3/2}}  
\right. \right. \nonumber \\
& & \left.\left. + l \frac{\Gamma(l+1/2)}{(1+T_{\rm e}/T_{\rm 
s})^{l+1/2}} 
B_{\rm e} + \frac{l(l-1)}{2} 
\frac{\Gamma(l-1/2)}{(1+T_{\rm e}/T_{\rm s})^{l-1/2}} 
B^2_{\rm e} \right] \right\} . \nonumber
\end{eqnarray}
One way to derive the last expression is to use the identity 
$\sum_{k=0}^\infty (-1)^k\frac{\Gamma(k+r)}{\Gamma(k+1)}c^k = 
(1+c)^{-r}\Gamma(r)$, valid for any $c\in (-1,1)$. For $c=1$ 
the left hand side of the identity is not convergent, but we 
can still define it as $\Gamma(r)/2^r$, using the continuity 
property of the right hand side.

Using the above analytical expressions, we can calculate easily 
the quantity $^{(\pm)}F^{(\pm)}_l$ for any $l$. As mentioned in 
the beginning of this section, in the concrete calculations we 
refer to the case in which the superconductor is Al, with 
$\Delta \approx 200\ {\rm \mu eV}$. Then for a temperature 
$T_{\rm e} = 0.1$ K, $B_{\rm e} \approx 46$, so we shall neglect the 
terms $I_{4_l}$ and $I_{6_l}$, since they are proportional to 
$\exp{(-B_{\rm e})}$. Moreover, it should be noted that three of 
the other terms, $I_{2_l}$, $I_{3_l}$, and $I_{5_l}$, are all proportional 
to $\exp{(-A_{\rm s})}$. Among these three, $I_{5_l}$ contains 
the highest orders for $l=$ 1 and 2, while for $l=0$ all of them 
are of the same order. Therefore care should be taken to calculate 
consistently the highest order of the terms proportional to 
$\exp{(-A_{\rm s})}$.

\subsection{Power flux through the NIS junctions}

\begin{figure}[t]
\begin{center}
\unitlength1mm\begin{picture}(100,80)(0,0)
\put(0,0){\epsfig{file=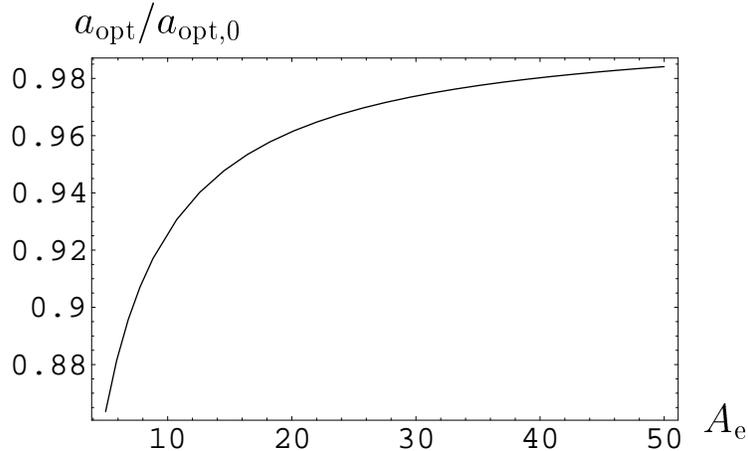,width=100mm}}
\end{picture}
\caption{The ratio between $a_{\rm opt}$, the value of $a_{\rm e}$ at 
maximum cooling power and finite temperature $T_{\rm e}$, and its limit 
value at zero temperature, $a_{\rm opt,0}$, as a function of 
$A_{\rm e} \equiv \Delta/k_{\rm B} T_{\rm e}$. Note 
that a range interesting for applications would be $A_{\rm e} > 20$.}
\label{A_max}
\end{center}
\end{figure}

In order to study the power flux through the NIS junctions, in 
the limit of low temperatures, let us take just the leading orders 
in $A_{\rm e}$, $A_{\rm s}$, $B_{\rm e}$, and $B_{\rm s}$ from the 
formulae given for $I_{j_l}$, $j=1, \ldots, 6$ and $l=1$. After 
dropping the irrelevant terms we arrive at 
the following expression \cite{teza}:
\begin{equation} \label{pe_aprox}
\dot{Q}_{\rm J} = \frac{\Delta^2}{e^2R_{\rm T}}
\sqrt{\frac{\pi}{2 A_{\rm e}^3}} \left[\frac{1}{2}g_{3/2}(a_{\rm e})
+ a_{\rm e}g_{1/2}(a_{\rm e})\right] 
- \frac{\Delta^2}{e^2R_{\rm T}} \sqrt{\frac{2\pi}{A_{\rm s}}} 
e^{-A_{\rm s}} \, .
\end{equation}
A problem that occurs in all the cooling experiments by NIS junctions 
is the heating of the superconductor. Due to the poor heat transport 
properties of the superconductor, the temperature $T_{\rm s}$ 
increases too much and 
$\dot{Q}_{\em J}$ becomes negative, or very close to zero, for any 
bias voltage. In such a case the cooler is practically ineffective. 
The heating 
can be controlled to some extent by placing normal metal traps on top of 
the superconductor. This has been discussed in Refs. \cite{trap,teza}. 
To avoid getting into too many details, here we shall consider 
$T_{\rm s}$ as an external parameter which is 
independent of the applied bias voltage (for any {\em specific} 
design more detailed calculations can be made). In case of 
an {\em ideal quasiparticle trap}, we set
$T_{\rm s} = T_2$. For a constant $T_{\rm e}$, $\dot{Q}_{\em J}$ is just 
a function of $a_{\rm e}$. This function reaches its maximum value 
at $a_{\rm e} = a_{\rm opt,0} \approx 0.66$ \cite{teza}, which 
corresponds to what we shall call the {\em optimum bias voltage}, 
$V_{\rm opt} \equiv \Delta - k_{\rm B} T_{\rm e} a_{\rm opt,0}$. From 
here we can draw an important conclusion: in the limit of low temperatures, 
the difference between 
the optimum bias voltage, $V_{\rm opt}$, and the energy gap $\Delta$ 
in the superconductor is independent of $\Delta$ and {\em scales with 
the temperature} of the ES in the normal metal. This means that 
$\delta_{\rm opt} \equiv \Delta - V_{\rm opt} \to 0$ as $T_{\rm e} \to 0$, 
and $a_{\rm opt}(T_{\rm e}) \to a_{\rm opt,0}$ as $T_{\rm e} \to 0$, 
where $a_{\rm opt}(T_{\rm e})$ is the exact value of $a_{\rm e}$ at 
the maximum cooling power, at finite $T_{\rm e}$. The ratio between 
$a_{\rm opt}$ and $a_{\rm opt,0}$ as a function of $A_{\rm e}$, is 
shown in Fig. \ref{A_max}. We have to note that the formulae containing 
$g_{k}(a_{\rm opt,0})$ can be simplified by replacing this function 
neither by $g_{k}(0) \equiv |(1-2^{1-k}) \zeta(k)|$ nor by 
$\exp{(-a_{\rm opt,0})}$ in the limit of low temperature. Replacing 
$a_{\rm e}$ by $a_{\rm opt,0}$ in Eq. (\ref{pe_aprox}), we obtain the 
optimum cooling power \cite{jukka}:
\begin{equation} \label{p_max}
\dot{Q}_{\rm J,opt} \approx 0.59\frac{\Delta^2}{e^2 R_{\rm T}}
\left(\frac{k_{\rm B}T_{\rm e}}{\Delta}\right)^{3/2}  
- \frac{\Delta^2}{e^2R_{\rm T}} \sqrt{\frac{2\pi 
k_{\rm B}T_{\rm s}}{\Delta}} e^{-\Delta/k_{\rm B}T_{\rm s}} \, .
\end{equation}

\section{Results}

Based on Eqs. (\ref{nis}) and (\ref{delta2p}) and using the formulae 
given in Sections \ref{n_terms} and \ref{analytic}, we can make 
concrete calculations of $NEP$ in the TSE of the microbolometer.
In Ref. \cite{DAJ} the direct and indirect cooling methods were 
compared, for the case of zero bias and optical input power. 
The conclusion was that $NEP$ was more than one order of magnitude 
lower in the case of indirect cooling than in the case 
of direct cooling, clearly recommending the first method for applications. 
When the optical input power is not zero the indirect cooling of the 
TSE might not be enough, due to the poor electron-phonon coupling at 
low temperatures [see Eq. (\ref{wellst})]. Therefore another 
set of junctions should be attached directly to the TSE to compensate 
for the optical power. To evaluate the situation, let us suppose 
that the resistances 
of these junctions are calibrated in such a way that 
$\dot{Q}_{\rm J,opt} = \dot Q_{\rm b} + \dot Q_{\rm oe}$ (we take 
$\dot Q_{\rm op} \equiv 0$), so $T_{\rm e}=T_1$. We consider 
that $T_1 = T_2 = 0.1$ K and that the superconductor is
in contact with an ideal quasiparticle trap ($T_{\rm s} = T_2$). 
The optical power incident on the detector is due to the cosmic 
background radiation and corresponds to a temperature 
$T_{\rm b} = 3$ K. As seen form Eqs. (\ref{q_oe}) and (\ref{n_oe}), 
both $\dot{Q}_{\rm oe}(\omega_{\rm o})$ and 
$\langle\delta^2 \dot{Q}_{\rm oe,shot}(\omega_{\rm o}) 
\rangle_\omega$ depend on the 
detector dimensions, emissivity, bandwidth and frequency of 
the absorbed radiation. To eliminate the first three of these parameters, 
we shall use the formula 
$
\langle\delta^2\dot{Q}_{\rm oe,shot}(\omega_{\rm o})\rangle_\omega  
= 2\hbar\omega_{\rm o}\dot{Q}_{\rm oe}(\omega_{\rm o})
$
and treat $\dot{Q}_{\rm oe}(\omega_{\rm oe})$ as a parameter obtainable 
from experiment. On the other hand, making use of the formulae 
given for the functions $I_{p_2}$, $p = 1, \ldots, 6$ and 
keeping only the highest orders in $A_{\rm e} = A_{\rm s}$, we 
obtain the following expression for the fluctuations:
\begin{equation} \label{J_noise}
\langle\delta^2 \dot{Q}_{\rm J,shot}\rangle_\omega \approx 
2.05 \frac{\Delta^3}{e^2 R_{\rm T}} 
\left(\frac{k_{\rm B}T_{\rm e}}{\Delta}\right)^{5/2} + 
4\sqrt{2\pi} \frac{\Delta^3}{e^2 R_{\rm T}} 
\left(\frac{k_{\rm B}T_{\rm e}}{\Delta}\right)^{1/2} 
e^{-\Delta/k_{\rm B}T_{\rm e}} \, .
\end{equation} 
Now we can compare Eqs. (\ref{p_max}) and (\ref{J_noise}).
If we neglect the terms proportional to 
$e^{-\Delta/k_{\rm B}T_{\rm e}}$, we can write in general 
\begin{equation}
\langle\delta^2 \dot{Q}_{\rm J,shot}\rangle_\omega \approx 
3.47 k_{\rm B} T_{\rm e} \dot{Q}_{\rm J,opt}(\omega_{\rm o}) = 
3.47 k_{\rm B} T_{\rm e} \dot{Q}_{\rm oe}(\omega_{\rm o}) \, . \nonumber
\end{equation}
Comparing the above equation with the equation for 
$\langle\delta^2 \dot{Q}_{\rm oe,shot}(\omega_{\rm o}) 
\rangle_\omega$, we obtain
$\langle\delta^2 \dot{Q}_{\rm J,shot}\rangle_\omega \approx 
1.74 (k_{\rm B}T_{\rm e}/\hbar\omega) 
\langle\delta^2 \dot{Q}_{\rm oe,shot}(\omega_{\rm o})\rangle_\omega$. 
If we write $\omega_{\rm o} \equiv x \omega_{\rm o,max}$, where 
$\omega_{\rm o,max}$ is the angular frequency corresponding to 
the maximum of the energy density of the cosmic background radiation, 
$\hbar\omega_{\rm o,max} = y k_{\rm B} T_{\rm b}$ ($y \approx 2.82$), 
then we can write 
\begin{equation} \label{comparatie}
\langle\delta^2 \dot{Q}_{\rm J,shot}\rangle_\omega \approx 
0.62 \frac{T_{\rm e}}{x T_{\rm b}} 
\langle\delta^2 \dot{Q}_{\rm oe,shot}(x)\rangle_\omega \, .
\end{equation}
Since $\partial\dot{Q}_{\rm J}/\partial V = 0$ 
at the optimum bias voltage, Eq. (\ref{upsi}) 
simplifies and we obtain $\Upsilon(\omega) = 
\langle\delta^2\dot{Q}_{\rm J,shot}\rangle_\omega$. 
In experiments it is expected that 
$T_{\rm b}/T_{\rm e} \ge 30$, so, for $x$ of the order of 1 
[Infra Red (IR) radiation] 
the shot noise in the junction, in the case of the combined cooling 
method, is very small as compared to the shot noise due to the background 
radiation. On the other hand it should be noted that the optical power 
of the IR radiation is absorbed by an 
antenna and is transformed into heat by the Joule effect in the 
TSE. In such a case, the noise contribution due to the input optical 
power may not be well described by Eq. (\ref{n_oe}) (here further 
investigation is needed). Therefore we shall write in general 
$
\langle\delta^2\dot{Q}_{\rm oe}(\omega_{\rm o})\rangle_\omega  
= 2\phi \hbar\omega_{\rm o}\dot{Q}_{\rm oe}(\omega_{\rm o})
$, 
where $\phi$ is a parameter of value between 0 and 1.

Taking again the two limit situations from Ref. 
\cite{DAJ}, and using Eq. (\ref{comparatie}) we say that if 
the Kapitza resistance is much 
smaller than the thermal resistance between the electrons 
and phonons, then 
\begin{eqnarray}
NEP^2&\approx& \langle\delta^2 \dot{Q}_{\rm ep,shot}\rangle_\omega 
+\langle\delta^2 \dot{Q}_{\rm K,shot}\rangle_\omega
\left(\frac{\partial \dot{Q}_{\rm ep}}{\partial T_1}\right)^2
\left(\frac{\partial \dot{Q}_{\rm K}}{\partial T_1}\right)^{-2} \nonumber \\
& & +\left(\phi + 0.62 \frac{T_{\rm e}}{x T_{\rm b}} \right) 
x y k_{\rm B} T_{\rm b} \dot{Q}_{\rm oe}(x) \nonumber \\
&\approx& \langle\delta^2 \dot{Q}_{\rm ep,shot}\rangle_\omega 
+ \left(\phi + 0.62 \frac{T_{\rm e}}{x T_{\rm b}} \right) 
x y k_{\rm B} T_{\rm b} \dot{Q}_{\rm oe}(x) ,  \label{nep1}
\end{eqnarray}
where the last approximation holds if $(T_2/T_1)^5$ is much smaller 
than the ratio between the electron-phonon and the Kapitza 
resistance [$(T_2/T_1)^5(\Sigma \Omega T_1/\Sigma_{\rm K}S)\ll 
1$], which is certainly the case for the indirect and combined cooling, 
under the assumptions made. 
In the other limiting situation, when the Kapitza resistance is much 
higher than the thermal resistance between the electrons 
and phonons, we can write 
\begin{equation}\label{nep2}
NEP^2\approx 4\langle\delta^2 
\dot{Q}_{\rm ep,shot}\rangle_\omega 
+ \left(\phi + 0.62 \frac{T_{\rm e}}{x T_{\rm b}} \right) 
x y k_{\rm B} T_{\rm b} \dot{Q}_{\rm oe}(x) \, . 
\end{equation}
Note that the coupling between the phonon and electron systems 
introduces a factor 4 in front of $\langle\delta^2 
\dot{Q}_{\rm ep,shot}\rangle_\omega$ in this situation. 
As in \cite{DAJ} we assume 
that the volume of the TSE is $\Omega = 1\ {\rm \mu m^3}$ and if 
we take $T_1 = T_{\em e} = 0.1$ K, we obtain 
$\langle\delta^2 \dot{Q}_{\rm ep,shot}\rangle^{1/2}_\omega \approx 
7.43 \times 10^{-19} \ {\rm W/\sqrt{Hz}}$. Let us suppose that 
we detect IR radiation and suppose further that in such a case 
$\phi \approx 0$. From Eqs. (\ref{nep1}) 
and (\ref{nep2}) we can calculate two critical input optical powers, 
$\dot{Q}_{\rm oe,1}(x)$ and $\dot{Q}_{\rm oe,2}(x)$, respectively, 
by the equation 
$\dot{Q}_{\rm oe,1}(x) = \dot{Q}_{\rm oe,2}(x)/4 = 
\langle\delta^2 \dot{Q}_{\rm ep,shot}\rangle_\omega 
/(0.62 yk_{\rm B}T_{\rm e})$. For 
our choice of temperatures we obtain 
$\dot{Q}_{\rm oe,1}(x) \approx 2.29\times 10^{-13}$ W
and $\dot{Q}_{\rm oe,2}(x) \approx 9.15\times 10^{-13}$ W.
Therefore, for optical input power smaller than 
$\dot{Q}_{\rm oe,1/2}(x)$ (depending whether we are in the first or 
in the second case), the noise contribution comes essentially from the 
electron-phonon coupling. For higher input power, the noise due to 
the electron tunneling through the NIS junctions dominates. As seen 
from Eq. (\ref{comparatie}), in such a case the $NEP$ has a weaker 
dependence on temperature, being proportional to $T_{\rm e}^{1/2}$, 
for constant optical input power. 

Throughout this paper we did not refer to the {\em measurement} 
of the temperature of the ES of the TSE. As mentioned in the 
beginning, there exist different methods to do that and we 
want to keep the arguments here as general as possible. Therefore,
let us suppose that in an experiment the temperature is calculated 
from the measured value of a quantity $M(T_{\rm e})$ (for example $M$ can 
be the voltage across the NIS junction biased at constant current, 
or the current in a voltage biased transition-edge sensor). If the 
inaccuracy in reading the quantity $M$ is 
$\langle \delta_1 M\rangle_\omega$ (this should be significant just in 
badly designed experiments) and the total mean square 
fluctuations of $M$ (for example the shot noise added quadraticaly to 
the amplyfier noise) are  $\langle \delta^2 M_{\rm f} 
\rangle_\omega$, then the uncertainity in the measurement of $M$ is: 
\begin{equation} \label{nepf}
\langle \delta^2 M\rangle_\omega^{1/2} = 
\sqrt{\langle \delta^2 M_{\rm shot} \rangle_\omega + 
2\frac{\partial M}{\partial T_{\rm e}}  
Re \Big(\langle \delta M \delta T_{\rm e} \rangle_\omega \Big) +
\left( \frac{\partial M}{\partial T_{\rm e}} \right)^2
\langle\delta^2 T_{\rm e}\rangle_\omega} + 
\langle \delta_1 M\rangle_\omega \, ,
\end{equation}
where we assumed again that the changes in the chemical potential 
$\mu$, due to temperature fluctuations, are negligible and 
$Re (\langle \delta M \delta T_{\rm e} \rangle_\omega )$ 
represent the real part of the correlations between the fluctuations 
of $M$ and of $T_{\rm e}$. To evaluate Eq. (\ref{nepf}) we have to 
introduce in Eqs. (\ref{set}) the noise contribution due to the 
bias power. In this way we obtain extra 
terms in the expressions for $\Delta_T(\omega)$ given by Eqs. 
(\ref{delta1}) and (\ref{delta2}). Denoting the new function, 
influenced by the measurement process, by 
$\Delta_{{\rm m}_T}(\omega)$, and using in an obvious manner the two 
expressions for $\Delta_T(\omega)$, we write the equivalent 
of Eqs. (\ref{delta1}) and (\ref{delta2}) as 
\begin{equation} \label{delta3}
\Delta_{{\rm m}_T}(\omega) \equiv \delta T_{\rm e}(\omega) Z_{NEP_\omega}
= \Delta_T(\omega) + \delta T_{\rm e}(\omega) 
\frac{\partial\dot{Q}_{\rm b}}{\partial T_{\rm e}} 
\end{equation}
and 
\begin{equation} \label{delta4}
\Delta_{{\rm m}_T}(\omega) = \Delta_T(\omega) + 
\delta \dot{Q}_{\rm b,shot}(\omega)  - 
\frac{\partial\epsilon_{\rm F}}{\partial N}
\frac{\partial \dot{Q}_{\rm b}}{\partial E_{\rm b}}\frac{\delta
\dot{N}_{\rm b,shot}(\omega)}{i\omega} 
\end{equation}
respectively, where $E_{\rm b}$ is the bias voltage of the thermometer 
and $\dot{N}_{\rm b,shot}$ is the shot noise of the particle current 
due to the thermometer. The presence of the last term in Eq. 
(\ref{delta4}) depends on the temperature measurement setup. It 
should also be noted that when $\dot{Q}_{\rm b}$ is due to a flux 
of particles, so when $\dot{N}_{\rm b}$ is not zero, the $\pi/2$ angle 
between the phases of the noise spectra of these two quantities 
annihilates their correlations. Taking this into account, we can 
write in general
\[
\langle \Delta^2_{{\rm m}_T}(\omega) \rangle = 
\langle \Delta^2_T(\omega) \rangle + 
\langle\delta^2 \dot{Q}_{\rm b,shot}\rangle_\omega + 
\frac{1}{\omega^2}\left(\frac{\partial\epsilon_{\rm F}}{\partial N}
\frac{\partial \dot{Q}_{\rm b}}{\partial E_{\rm b}}\right)^2
\langle\delta^2\dot{N}_{\rm b,shot}\rangle_\omega 
\]
In this situation, 
the total Noise Equivalent Power, $NEP_{\rm t}$, should be defined as the 
input power that would produce the same change in the 
quantity $M$ as the one in Eq. (\ref{nepf}). 
Simple calculations lead us to the general result 
\begin{eqnarray}
NEP_{\rm t_\omega} &=& \sqrt{
\langle \Delta^2_{{\rm m}_T}(\omega) \rangle + 
2 \left( \frac{\partial M}{\partial T_{\rm e}} \right)^{-1} 
|Z_{NEP_\omega}|
Re \Big(\langle \delta M_{\rm f} \delta T_{\rm e} \rangle_\omega \Big) +
\left(\frac{\partial M}{\partial T_{\rm e}} \right)^{-2} 
|Z_{NEP_\omega}|^2\langle \delta^2 M_{\rm f} \rangle_\omega} \label{nept} \\
& & + \left|\frac{\partial M}{\partial T_{\rm e}} \right|^{-1} 
|Z_{NEP_\omega}|\langle \delta_1 M\rangle_\omega  \, , \nonumber
\end{eqnarray}
where $Z_{NEP_\omega}$ was defined in Eq. (\ref{delta3}). 
Using Eq. (\ref{delta4}) we can write
\[
Re \Big(\langle \delta M_{\rm f} \delta T_{\rm e} \rangle_\omega \Big) = 
Re \left[ \left\langle \delta M_{\rm f} \left(\delta \dot{Q}_{\rm b,shot}  - 
\frac{\partial\epsilon_{\rm F}}{\partial N}
\frac{\partial \dot{Q}_{\rm b}}{\partial E_{\rm b}}\frac{\delta
\dot{N}_{\rm b,shot}}{i\omega}\right) \right\rangle_\omega/ 
Z_{NEP_\omega} \right] \, .
\]
For the case when the thermometer is a NIS junction, current or voltage 
biased, the $NEP$ was calculated in Refs. 
\cite{golwala,golubev,private} in the limit of zero Kapitza resistance, 
infinitely good reading accuracy of quantity $M$, and when the 
only refrigerating junction in the system is the thermometer junction.

\section{Conclusions}

In this paper we calculated the temperature fluctuations and the 
Noise Equivalent Power $NEP$ in the thermal sensing element TSE 
of a microbolometer cooled by NIS junctions. 
From the general result we extracted simple expressions for two limiting 
cases. In the first case we considered that the Kapitza resistance is much 
smaller than the thermal resistance between the electrons 
and phonons, while in the second case we considered the opposite 
situation. In both cases the important role in the calculation of the 
$NEP$ was played by the electron-phonon shot noise and the 
noise due to the cooling NIS junction, coupled directly to the TSE, and 
to the input optical power [see Eqs. (\ref{nep1}) and (\ref{nep2})].

We gave analytical expressions for the quantitative evaluation 
of the processes that take place in NIS junctions, which 
are valid in the range of low temperatures. The validity 
of these formulae can be 
checked by the evaluation of the lower order terms in the quantities 
$A_{\rm s/e}$ (see the definitions in Section \ref{general}). 
Using the analytical expressions we showed that in the limit of low 
temperature of the electron gas 
in the normal metal ($T_{\rm e}$) the difference 
between the gap energy in the superconductor, $\Delta$, and the bias voltage 
corresponding to the maximum cooling power, $V_{\rm opt}$ (supposing that 
the temperature of the superconductor does not vary with the 
bias voltage), is independent of $\Delta$ 
and scales with $T_{\rm e}$, like $\Delta - V_{\rm opt} = 0.66 k_{\rm B} 
T_{\rm e}$. This is a very useful result since in such a case, in 
the quantitative evaluation 
of the processes that take place in NIS junctions, we can use neither 
the limit $(\Delta - V_{\rm opt})/k_{\rm B} T_{\rm e} \gg 1$, nor 
the limit $(\Delta - V_{\rm opt})/k_{\rm B} T_{\rm e} \ll 1$, for 
$T_{\rm e} \to 0$.

In the end 
we took into account in a general way the effect of the measurement 
process on the Noise Equivalent Power and we gave general expressions 
for its calculation. As expected, the measurement 
increases the uncertainty in the detected input optical power 
[see Eq. (\ref{nept})]. 

\section{Acknowledgements}

We thank A. Luukanen for many discussions. 
This work was supported by the Academy of Finland under the Finnish
Center of Excellence Programme 2000-2005 (Project No. 44875, Nuclear and
Condensed Matter Programme at JYFL).

\end{document}